\documentclass[letterpaper,notitlepage,fleqn]{article}

\usepackage{amsmath,amsfonts,amssymb} 
\usepackage{verbatim}
\usepackage[T1]{fontenc}        % används för att få svensk avstavning
\usepackage[english]{babel}     
\usepackage[latin1]{inputenc}   % och å, ä ,ö
\usepackage[dvips]{graphicx}
\usepackage{url}
\usepackage{ifthen}

\newcommand{\urlBiBTeX}[1]{\url{#1}}
\newcommand{\smallcaption}[1]{\caption{{\small #1}}}

\date{30th November 2007}                          %Ta bort kommentaren om du inte vill ha med datum.

\begin{document}
\pagenumbering{arabic}

%\ifthenelse{\equal{\condition}{true}}{ } { }

\title{The effect of time distribution shape on simulated epidemic models}

\author{Martin Camitz \and Åke Svensson}

\maketitle 

\abstract{By convention, and even more often, as an unintentional consequence of design, time distributions of latency and infectious durations in stochastic epidemic simulations are often exponential. The skewed distribtion typically leads to unrealistically short times. We examine the effects of altering the distribution latency and infectious times by comparing the key results after simulation with exponential and gamma distributions in a homogeneous mixing model aswell as a model with regional divisions connected by a travel intensity matrix. We show a delay in spread with more realistic latency times and offer an explanation of the effect.

}

\newpage

\section{Introduction}

Exponential distributions for latency times and infectiousness times often appear with models of infectious diseases, simulated or solved analytically. The distribution does not resemble observed distribution of latency or infectious times. Depending on the problem at hand, this may be a reasonable simplification. For certain questions where the speed of spread of the infection is of less importance, this assumption may give perfectly satisfactory results. Recently research interest, however, has been directed in the initial highly random phase of the epidemic, whereas the final size of the epidemic is perhaps of less interest \cite{tommi, daley}. In spite of this the exponential time assumption has become off-the-wall and many authors, by tradition, disregard the consequence of their assumption.

The reason for the wide spread use is that the exponential distribution is inherently "memoryless" \cite{markov} which means that future predictions of the state of the epidemic in terms of number of latent and infectious individuals etc is based solely on the current state and not on the history of states. The probability that 10 people will have fallen ill on Friday depends only on how many are ill on Thursday. The state on Wednesday or Tuesday is irrelevant. This makes possible a simple stochastic simulation by utilizing a Markov process.

Exponential distributions will appear as a consequence of the assumption that the rate at which individuals leave a certain state at a certain time only depdends on how many individuals is in that state at this time. This corresponds to a constant hazard for any individual to leave the state is the same as the "memoryless" property. Many authors therefor include the exponential distribution assumption more of less unintentionally while design the model.

In this paper we show that the time distribution is vital for accurate results and also that, without abandoning a Markov approach, we are given some freedom to adapt the distribution to fit real data, using a gamma distribution.

Much work has been done to show the effect of traveling and migration on the evolution of epidemics \cite{rvachev, hufnagel, colizza1, colizza2, cooper}. For today's global outbreaks, notably the SARS outbreak of 2001, the need to incorporate what information we have on travel networks in our simulations has become increasingly apparent. Models that take the Markov approach seem well suited for this purpose which was demonstrated by Hufnagel et al. The population is divided into a number of local regions which can be countries, municipalities or other geographic or even social groupings. They are interconnected by an infectiousness intensity matrix describing how infection is tranferred between regions. This matrix can be estimated from, for example, travel data. 

Hufnagel et al. used the catchment area around each international airport and within these used a \emph{SEIR-model}, where every individual can be in one of the states \emph{susceptible}, \emph{latent}, \emph{infectious} and \emph{recovered}. These local processes were linked together by the network of international avaiation enabling the disease to be transmitted along flight paths.  

Camitz and Liljeros \cite{camitz} constructed a similar model of a SARS-like outbreak in Sweden. In this model the municipal borders were used to partition the country. Using detailed travel data between municipalities, a travel intensity matrix was estimated and the geograpic spread could be studied aswell as the effect of travel restrictions.

In more detail, the SLIR-model works as follows. The population in each municipality is assigned to one of four states, decribing their disease state: Susceptible, Latent, Infectious and Recovered. A susceptible may become latent with a probability which depends on the number of infectious, in his/her own aswell as connected municipalities, depending on the intensity of travel between connected municipalities. After being infected, the latent individual moves through stages L and R in times corresponding to known latency and infectious times. The actual time for an individual will vary randomly about the mean time, which is fixed. The crucial point is how these times vary. In \cite{hufnagel, camitz} the times are picked from an exponential distribution. 

\begin{equation*} \label{eq:stateflow}
S\rightarrow L\rightarrow I\rightarrow R
\end{equation*}

Indeed we are certainly not the first to introduce the Gamma distribution in these contexts. Gamma distributions have for a long time been $stadard$ in modeling progress of chronic diseases (e.g. cancer) through different stages. They have also been used in models of epidemics, see \cite{eichner} for a recent example. But discussion about using this and other distributions is lacking in research today. Times with single point distributions are sometimes considered a reasonable approximation \cite{carrat,kaplan} but for following the complete dynamics we feel that a variance is necessary. Other time distributions have also been used, such as uniform, Log-normal or Weibull, the latter two notably differing from Gamma primarily in their tails. Such distriutions may be appropriate but wull not be possible to model with the Markov approach.

\subsection{The exponential and gamma distribution}

\begin{figure}[p]
\begin{center}
\includegraphics[scale=.5]{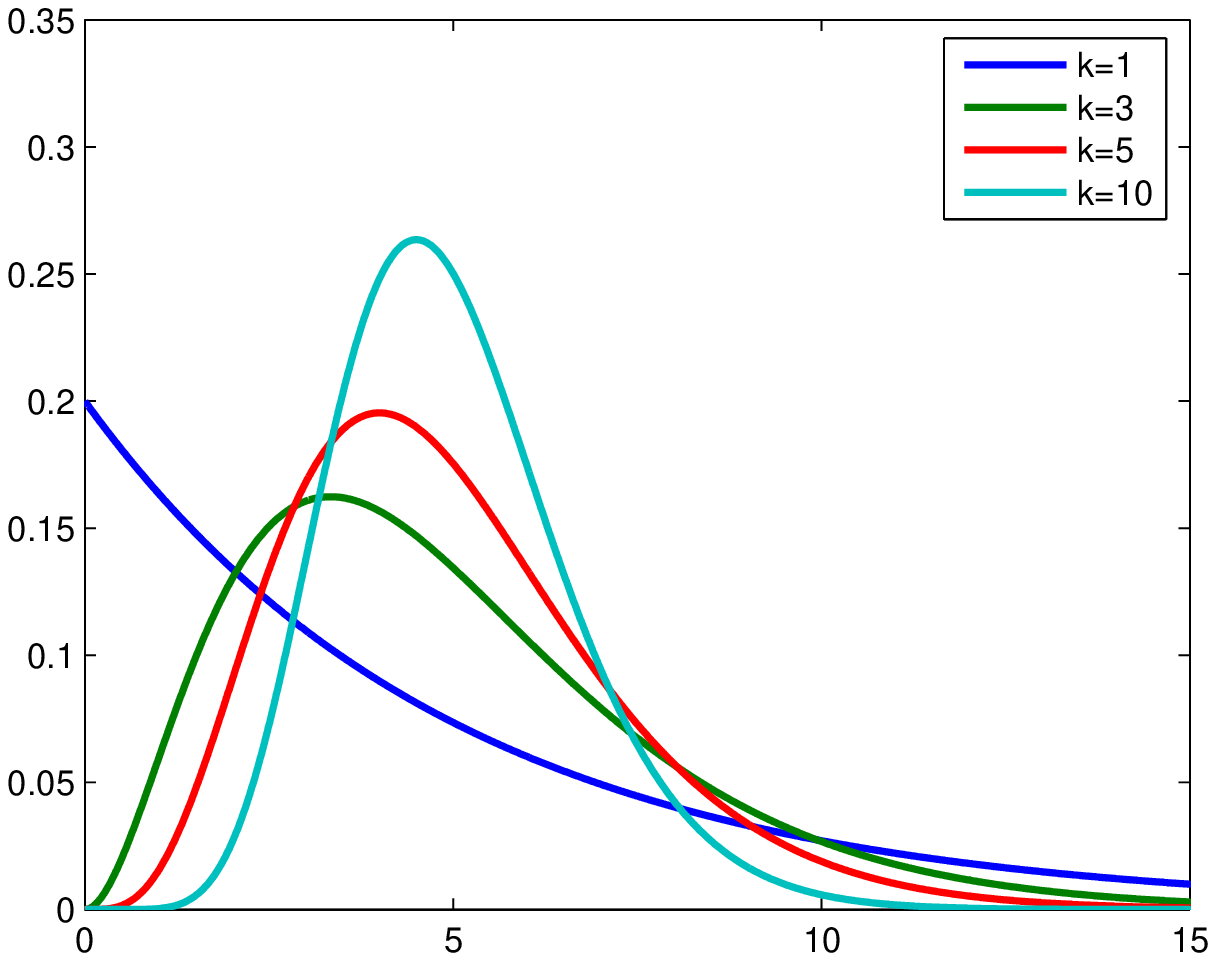}
\end{center}
\smallcaption{Probability distribution of the Gamma distribution for varying $k$, all with expectation value $5$. The special case of $k=1$ produces an exponential distribution. \label{fig:gamma_distr}}
\end{figure}

The main drawback with exponential times is a questionable tie to reality. The exponential distribution is highly skewed, with high densities for short times and a long tail. Empirical latency times and infectious times are not exponentially distributed, but rather have a symmetric density about their expectation values. Furthurmore, the exponential distribution has a quite high variance, equal to its expectation value squared, whereas empirical times tend to deviate little from the mean. The dark blue curve in \ref{fig:gamma_distr} shows a plot of the probability density function of the exponential distribution as a special case of the gamma distribution, the circumstances for this relationship to be explained later. The exponential distribution has a single parameter equal to the inverse of the expectation value.

In practical simulations this will shorten the time of interest. Since the median is lower than the expectation value implying that most times will be shorter than the expected. For example, say the expectation value of the latency time is set to 5 days. With an exponential distribution, 63\% of the random times will be shorter than 5 days. 18\% will be shorter than 1 day. Such short times are clearly unrealistic and furhtermore, latency times are expected to fall symmetrically about the mean. 

The additional disadvantage in stochastic epidemic simulations is that the outcome is highly dependent on the initial stages. Individuals with short latency times will predominantly make up the initially infected and will inevitably speed up the outbreak. That is to say that the skewness of the exponential distribution is dominant in the early stages of the simulation whereas the expectation value is not apparent until the stochasticity has averaged out. 

A few authors have proposed that the gamma distribution be used instead[ref]. The gamma distribution, denoted $\Gamma(\kappa, \theta)$ has two parameters, a shape parameter $\kappa$ and a scale parameter $\theta$. For integer $\kappa$:s the probability density function takes on a particularly simple form:

\begin{equation}
f(t;\kappa,\theta)=t^{\kappa-1}\frac{e^{-t/\theta}}{\theta^\kappa (\kappa-1)!}
\end{equation}

The mean is $\kappa \theta$ and variance $\kappa \theta^2$. For $\kappa =1$ the gamma distribution is in fact identical to the exponential distribution. Keeping the expectation value constant, with larger $\kappa $s, the gamma distribution becomes increasingly symmetric about its mean and start to resemble times distributions we have learnt to expect. The skewness of the density function is infact $2/\sqrt{\kappa}$.

The gamma distribution can actually be realized with an uncomplicated extension of a Markov model such as the one in \cite{camitz}. The sum of several exponentially distributed times is in fact gamma-distributed. This is expressed as follows. Let $X_1\ldots X_n$ be independant stochastic variables from an exponential distribution $\mathrm{Exp}(\xi)$. Then, $Y=\sum_{i=1}^n X_i$ belongs to $\Gamma(n, \xi)$.

In practice, instead of having only a single latency stage and a single infectious stage, we add stages, forcing each individual to go through several stages of latency before becoming infectious, and in the same manner, several stages of infectiousness before recovering. Thereby we achieve an arbitrarily symmetic time distribution with a minimal alteration to our SLIR-model. In doing so we alter $\kappa $ which is as shown is equivalent to the number of stages. These stages have no epidemiological meaning but serve only to change the appearance of the time-distribution.

At the same time we have to decrease $\xi$ so as to keep constant the expected time which is $n\xi$ in the gamma distribution. We can select any $\kappa $ we like to produce a good enough fit to an empirical distribution, or at the very least, the mean and variance. The cost of added stages is of course memory requirements but happily the simulation time is not influenced to a degree to be a deterent in any way. This is due to one of the key advantages of the Markov approach. We do not have to keep track of any individuals in the model. We simply record their number in eash state.

Using a modified version of \cite{camitz} we show that ignoring the shape of the time distribution devalues the results, comparing the results for different $\kappa$ for both latency times and infectious times. The difference in absolute terms is significant.

\section{Data and Methods}

We carried out two sets of four simulations, each consisting of 1000 realizations of an outbreak inititated with one infected individual in Stockholm. In the first set we confined the population of Stockholm allowing us to test the change employing the gamma distribution in a single locality random-mixing situation not complicated by travel. There is no specific reason for using Stochkolm either as the origin of infection or as a comfinement. The mixing model is the same in all municipalities. 

In the second set, we used the full travel network for a full scale simulation. In each set we ran a reference simulation with both the latency and infectious times distributed according to an exponential distribution with the mean 5 days. Except for different parameters, this setup corresponds exactly to the one used in [camitz]. The other three had gamma-distributed latency times, infectious times or both. In the case of gamma distributions, $\kappa=3$ was used. $\xi$ was adjusted to attain an expected time of, again, 5 days. 

The inter-municipal infectioussness matrix is the same as in Camitz \& Liljeros \cite{camitz}. It is based on an interview survey conducted in Sweden between 1999 and 2001 containing some 35000 journeys. This resulted in approximately 12000 matrix elements $\gamma_{ij}$ each estimated with 
\begin{equation}
\gamma_{ij}=\gamma M_{ij}/\sum_j M_{ij}
\end{equation}
 where $M_{ij}$ is the number of journeys per day from municipality $i$ to $j$ and $\gamma$ is a global scalar \cite{hufnagel}. 

The disease is a fictive moderatly infectious disease with an $R_0$ of $2.5$, within every homogenous subpopulation. 

To describe the state of the epidemic we introduce the vector $\mathrm{S}$ to keep track of the number of susceptibles in each municipality. Additionally, two sets of vectors ${\mathrm{L_1}\ldots \mathrm{L_{\kappa}}}$ and ${\mathrm{I}_1\ldots \mathrm{I_{\lambda}}}$ are defined to keep track of latents and infectious. The indexes $\kappa$ and $\lambda$ are the chosen first paramenters for the gamma distribution for latency and infectious times respectively. We will use a general formalism for the time being but later we set the parameters to either 1 or 3. In the first case, corresponding to an exponential distribution, there will only be one vector in the set. If $\kappa$ is greater than unity, then this will be the number of stages of latency or infectioussness that each individual needs to pass through. The sizes of each vector is of course equal to the number of municipalities. Let $P$ be this number. The dimensionality of the entire state space is equal to $D=P\cdot(1+\kappa+\lambda)$. The vectors are indexed as $I_{k,i}$ (italisized when indexed with $i$) where $i$ is the municipality and $k$ is the stage of disease. For any purposes they can be treated as tensors or matrices. Summing over all $k$s and $i$s yields in this case the total number of infected. Note that recording the number of recovered individuals is redundant since it is simply the sum of the number in the three states of infectiousness already covered, subtracted from the population.

At the start of the run the element $S_i$ of $S$ is equal to the population sizes $N_i$ of each municipality. This is the initial state in each run. For each municipality we now have $1+\kappa+\lambda$ possible state transitions, each involving incrementing an element corresponding to the municipality in one vector and decrementing the "preceding". This is true for all transitions except from the last stage of infectiousness which of course only involves a decrement.

We are now ready to set up the equations that will define the transition matrix of our Markov process. The quantities $Q_{ik}^\mathcal{X}$ below, is for each municipality $i$ the \emph{intensity} of individuals passing on to the next stage of illness and are connected to the probabilities of the corresponding state transitions. $\mathcal{X} \in \{\mathcal{L}_1\ldots \mathcal{L}_{\kappa}, \mathcal{I}_1\ldots \mathcal{I}_{\lambda}, \mathcal{R}\}$ is a \emph{label} signifying transitions to one of the latency states, one of the infectious states or the recovered state. It is written in a calligraphic font to avoid confusion with $\mathrm{L_k}$, $\mathrm{I_k}$ and $\mathrm{R}$ which are vectors.

\begin{eqnarray} \label{eq:intensities1}
Q_i^{\mathcal{L}_2} &=& \upsilon \kappa L_{1,i} \nonumber\\
&\vdots\nonumber\\
Q_i^{\mathcal{L}_{\kappa}} &=& \upsilon \kappa L_{\kappa-1,i} \nonumber\\
Q_i^{\mathcal{I}_1} &=& \upsilon \kappa L_{\kappa,i} \nonumber\\
Q_i^{\mathcal{I}_2} &=& \beta \lambda I_{1,i} \\
&\vdots\nonumber\\
Q_i^{\mathcal{I}_{\lambda}} &=& \beta \lambda I_{\lambda-1,i} \nonumber\\
Q^\mathcal{R} &=& \beta \lambda I_{\lambda,i} \nonumber
\end{eqnarray}
Finally, people are infected (become latent) with the intensity that depends number of infected in all the municipalities and the travelintensity between each of them:
\begin{equation}\label{eq:intensities2}
Q_i^{\mathcal{L}_1} = \left[\alpha \sum_{k=1}^{\kappa} I_{ki}  + \sum_{\begin{subarray}{1} j=1\\j\neq i \end{subarray}}^N \gamma_{ij} \sum_{k=1}^{\lambda} I_{kj} \right] \frac{S_i}{N_i}.
\end{equation}

In the equations above, $\alpha$ is the the expected number of secondary infected per infectious. $\upsilon$ is the inverse latency period and $\beta$ the recovery rate. The second row reads: The number of people per unit time leaving the first latency stage is the number of people in that stage times the number of stages times the scalar rate $\upsilon$. The last row is similar, as is the first term of the first row but summed over all infectious stages and also includes a factor to account for a decreasing number of susceptibles. The second term is the contribution from other municipalities via the infectiousness network. It includes a sum of infectious individuals over all stages and all municipalities but the current.

Each of these intensitities is the parameter required to specify the exponential distribution that yields the timesteps for the corresponding transition. The model is now in all respects in place. To simulate we would like to take each transition in order and so we are interested to know the time $\Delta t$ untill the next transition, given the current state. The time, one can easily show, is incidentally also exponentially distributed with parameter $Q$ equal to the sum of the $D$ intensities in Eqs. \ref{eq:intensities1} and \ref{eq:intensities2},

\begin{flushleft}
\begin{equation}
\Delta t \in \mathrm{Exp}(Q), 
\end{equation}
\end{flushleft}
\begin{equation}
Q = \sum_{i=1}^M (Q_i^{\mathcal{L}_1} + \cdots + Q_i^{\mathcal{L}_\kappa} + Q_i^{\mathcal{I}_k} + \cdots + Q_i^{\mathcal{I}_\lambda} + Q_i^{\mathcal{R}}).
\end{equation}

To determine which transition occurs at this time we compare the intensities among themselves. The probability of a transition is proportional to the relative value of the corresponding intensity, simply the intensity normalized by $Q$. So in each pass through the main loop of the algorithm we find $Q$, pick a random time step from the exponential distribution specified by $Q$ as a parameter, randomly pick a transition according to the relative value of the intensities, update the state vectors and the intensities according to the new state and start again. The simulation proceeds this way until an arbitrarily chosen time limit is reached or until there are no more infectious or latent, which ever comes first. In our case we chose 60 days as by this time a substantial majority of simulated scenarios will have developed into epidemics. Recall that the object of interest in not the final size but any delay in time of the epidemics.

\section{Results}

The prevalence, along with some additional results, from the first set of four simulations confined to Stockholm is presented in table \ref{table:sthlm}. It clear that the shape of the time-distribution determines the outcome of the simulated epidemic. What is more, the prevalence after 60 days follows the anticipated pattern, decreasing with more realistic latency times and increasing with more realistic infectious times.

\begin{figure}[p]
\begin{center}
\includegraphics[scale=1.0]{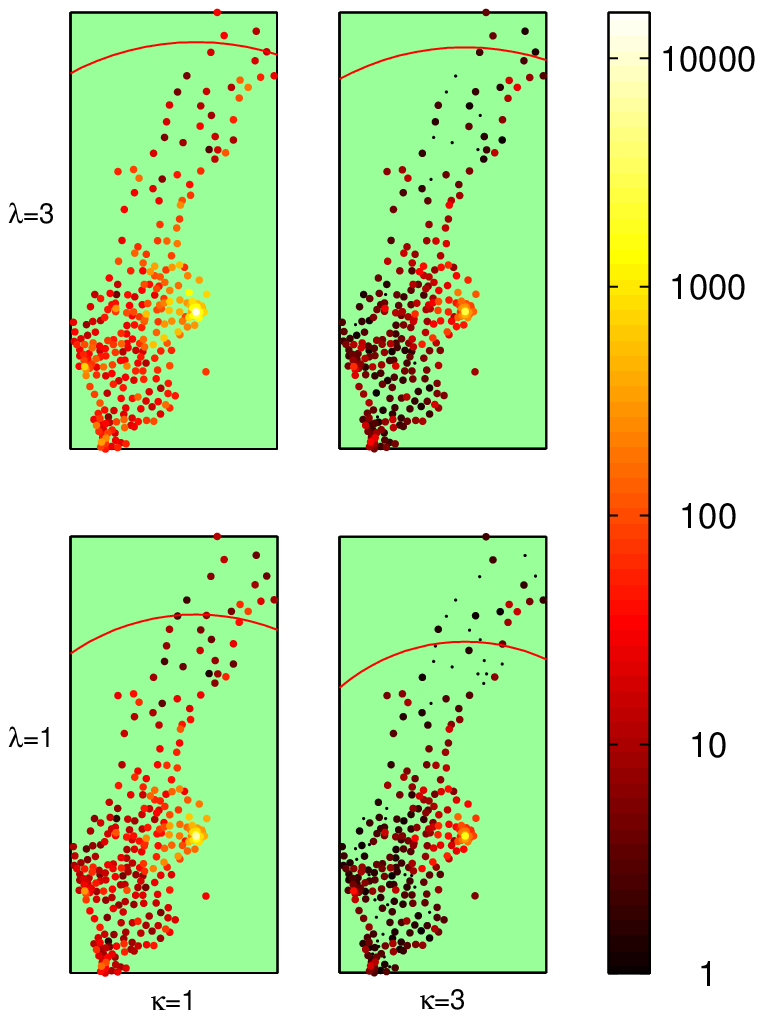}
\end{center}
\smallcaption{Image visualizing the epidemic state after 60 days simulation, averaged over 1000 runs. The form parameter for latency times increase from the left to right column and for infectious times from the bottom to top row. The prevelence in each municpality is color coded on a logarithmic scale. Clearly a more realistic latency time distribution delays the epidemic significantly.\label{fig:meanI}}
\end{figure}

The results of the second simulation set is presented in figure \ref{fig:meanI} as a geographic plot over Sweden with each municipality represented by a colored dot. The prevalence is represented by color on a logarithmic scale. Again, the incidence and geographic spread is highly dependent on the shape of the time-distribution, see also \ref{table:country}. As with the first set, the order of severeties after 60 days is the anticipated but th effects are even more apparent. Retransmission from connected municipalities amplifies the distribution effects.

\begin{figure}[p]
\begin{center}
\includegraphics[scale=.5]{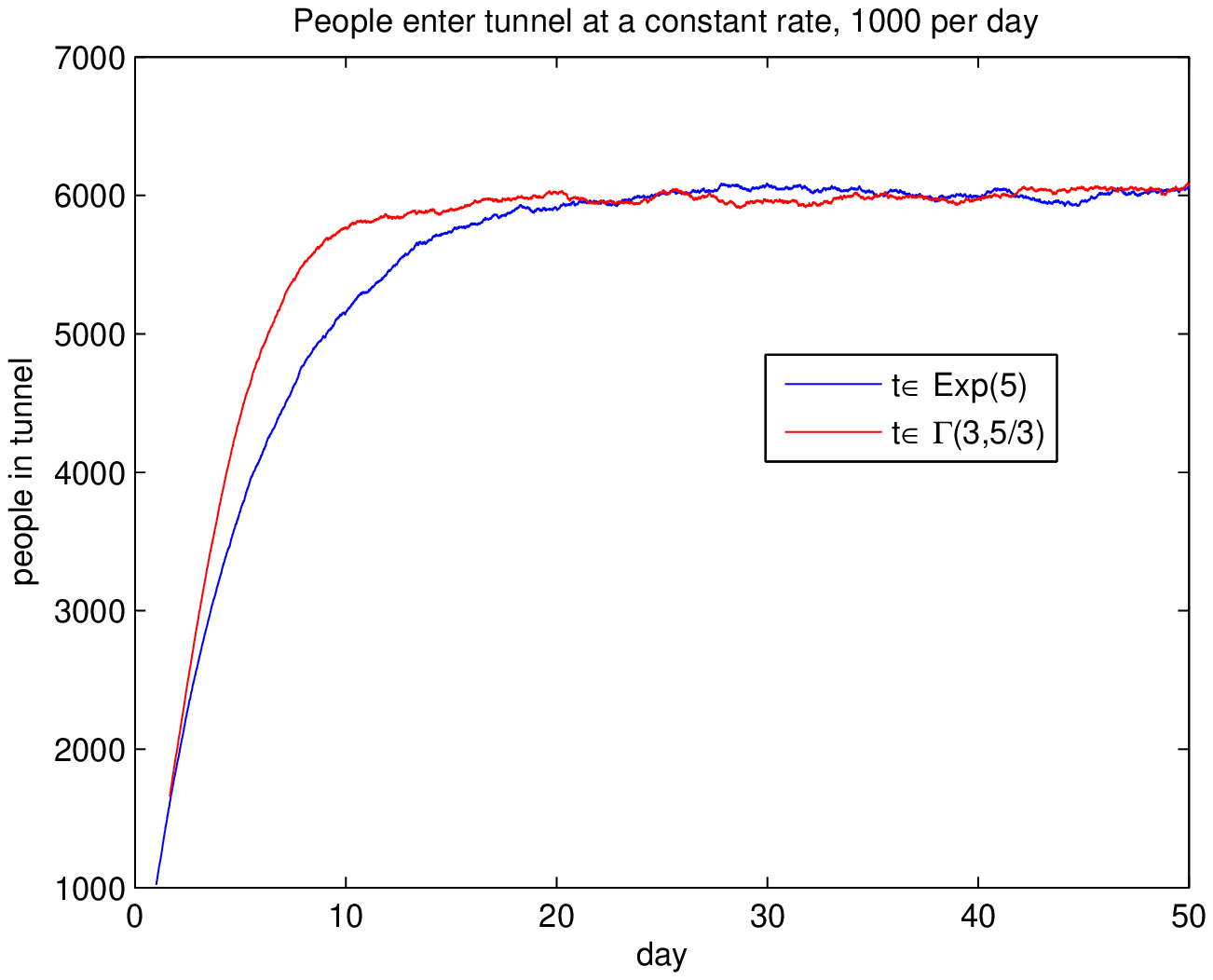}
\end{center}
\smallcaption{A simulation of people entering a cave at a rate 1000 per day at speeds selected from, in the case of the blue curve, an exponential distribution and in the case of the red curve, a gamma distribution with $k=3$. The number of people simultaneously in the cave is plotted. The expected passage time for both curve is 5 days which gives the same number of people in the cave after a transitional phase. The transitional phase differs, however. \label{fig:cave1}}
\end{figure}

Remember that there is a time limit of 60 days and that different latency time distributions do not necessarily affect the height of the incidence peak, only when it occurs. We also added a figure for additional support with $k$ simultaneously varied from $1$ to $20$.

\begin{table}[p]
\centering

\begin{tabular}{|lc|cc|}
\hline
&	 &	$\kappa$ &	 \\
 &	$\lambda$ &	1 &	3 \\
\hline
Cumulative incidence  &	3 &	10099 &	4174 \\
&	1 &	3816 &	1968 \\
\hline
Prevalence  &	3 &	2751 &	1021 \\
&	1 &	887 &	419 \\
\hline
Mean time for extinction (days)  &	3 &	5.4 &	5.6 \\
&	1 &	5.3 &	4.5 \\
\hline
Number of extinction runs  &	3 &	230 &	217 \\
&	1 &	359 &	390 \\
\hline

\end{tabular}
\smallcaption{Results for epidemic confined to Stockholm, essentially a homogeneous dispersion model. The figures follow the predicted behaviour. Note the differences in monoticity in extinction runs and mean time for extinction. We attribute differences in extinction runs and and mean time for extinction across rows (equal $\lambda$ to random variance and as an effect of the cut-off time, as they should theoretically be equal. \label{table:sthlm}}
\end{table}

\begin{table}[p]
\centering

\begin{tabular}{|lc|cc|}
\hline
&	 &	$\kappa$ &	$\lambda$ \\
 &	$\lambda$ &	1 &	3\\
\hline
Cumulative incidence  &	1 &	718830 &	140530\\
&	3 &	184240 &	44806\\
\hline
Prevalence  &	1 &	212600 &	35263\\
&	3 &	46341 &	9828\\
\hline
Mean incidence in municipalitites  &	1 &	736 &	122\\
&	3 &	160 &	34\\
\hline
Mean time for extinction (days)  &	1 &	4.4 &	3.9\\
&	3 &	3.5 &	3.3\\
\hline
Number of afflicted municipalities  &	1 &	279.3 &	250.7\\
&	3 &	249.0 &	190.6\\
\hline
Number of extinction runs  &	1 &	95 &	99\\
&	3 &	241 &	295\\
\hline
Fraction infected from &	1 &	71  &	72 \\
another municipality (\%) &	3 &	33  &	33 \\
\hline

\end{tabular}

\smallcaption{The results for the full simulations over all municipalities. The behaviour exhibited in the single municipality simulations is even more apparent here which means that retransmission from connected municipalities amplifies the distribution effects.\label{table:country}}
\end{table}

\section{Discussion}

Considering first the simpler case of a single municipality, the extremely short latency times generated by the exponential distribution was expected to accelerate the epidemic. More individuals become infectious early in the simulation, in turn infecting others earlier. It can be shown, however, that with shorter mean latency time the final size of the epidemic is unchanged. With a skewed infectious time distribution the effect is reversed. The epidemic will be delayed, at least initially, due to the abundance of very short infectious times. Each infected will infect a fewer number of secondary infecteds before recovering. It is harder for the epidemic to catch on and the probability of the disease dying out completely is higher. 

\begin{figure}[p]
\begin{center}
\includegraphics[scale=.5]{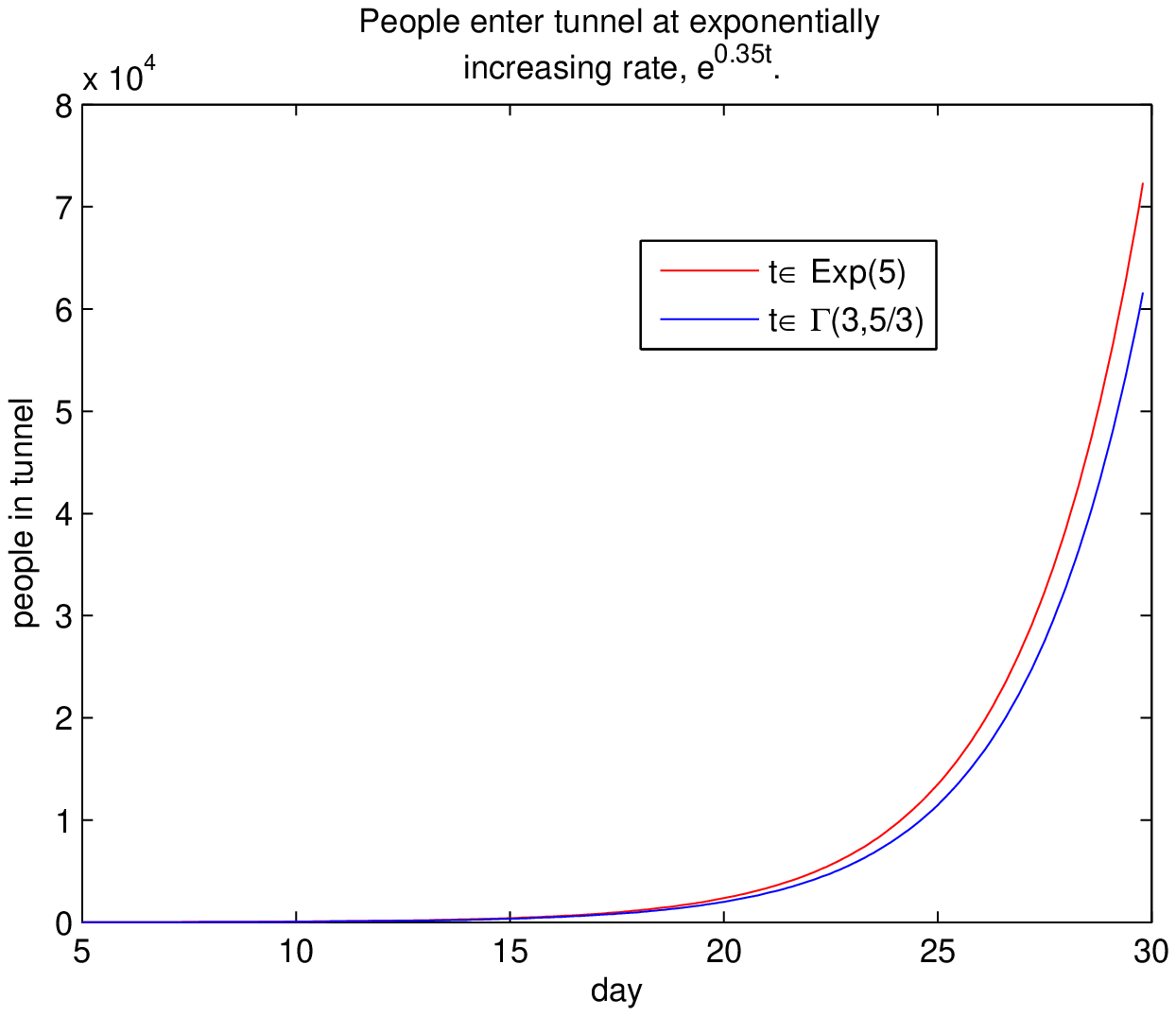}
\end{center}
\smallcaption{Analogous plot as figure \ref{fig:cave1} but with exponentially increasing entrance rate, more suited to portraying epidemic growth. As long as the rate increases this way, there will never be a steady state where the number of people in the cave for the two cases are equal.\label{fig:cave2}}
\end{figure}

Although this may not be immediately apparent, longer times also means more individuals in the different stages. Figures \ref{fig:cave1} and \ref{fig:cave2} exemplify this. Here we've simulated people walking through a tunnel. As people emerge from the tunnel we count the number still inside. Everybody walks at different speeds. The time it takes for them to get to the other side is random, on average 5 days, but in one case (red curve) the time is taken from a gamma distribution and in the other, blue curve, an exponential distribution. In the first graph, people enter at a constant rate of 1000 per day. As can easily be visualized, after a while a steady state is reached where both distributions give rise to the same number of people inside the tunnel. Afterall, the average time is 5 days in both cases - in the steady state, as many should exit as enter the cave, 1000 per day, regardless of the distributions. What is more interessting is before the steady state is reached. Here the high number of speeders in the exponential case clearly make their mark in the statistics, quickly exiting the cave and leaving a fewer number left inside. Only after the steady state has been are the slow-walkers inside sufficiently numbered to make up for the speeders.

\begin{figure}[p]
\begin{flushleft}
\includegraphics[scale=.8]{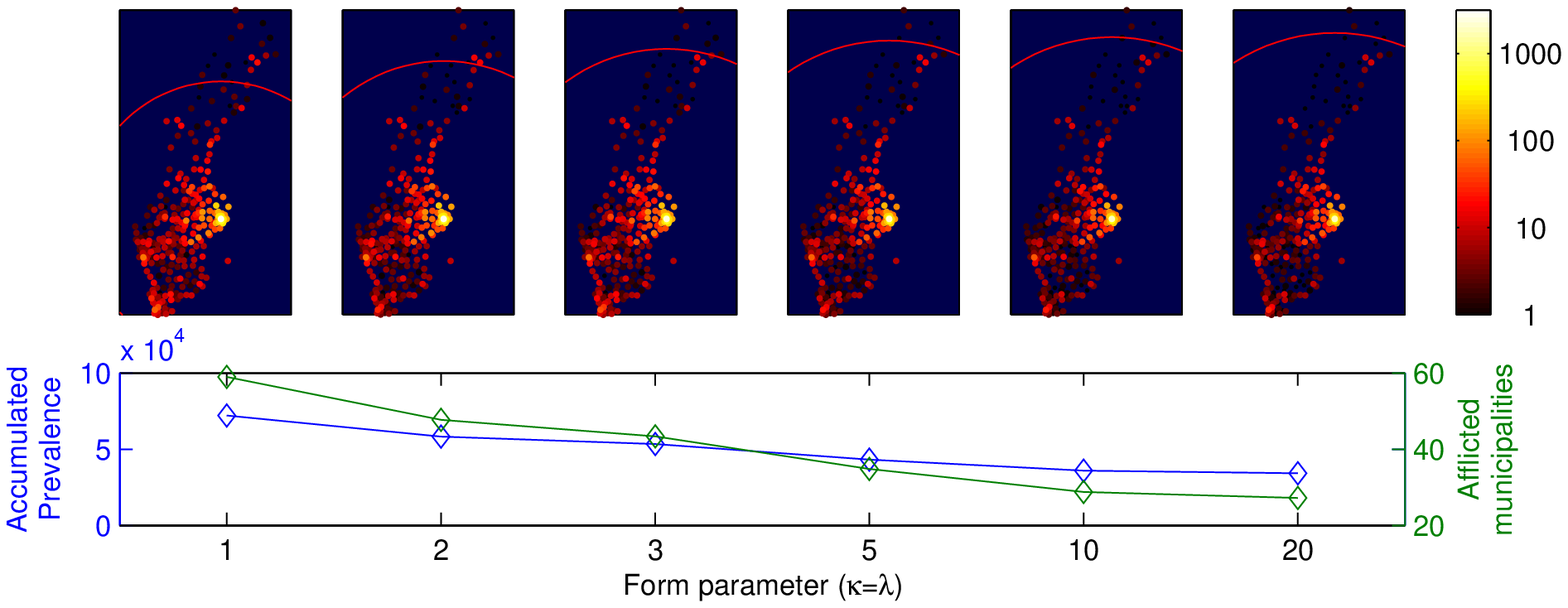}
\end{flushleft}
\smallcaption{Here the form parameter for both latant and time distributions are set equal. Cumulative incidence i.e. the total number of infected, is plotted below for each setting. \label{fig:meanI_range}}
\end{figure}

Since we are dealing with stochastic simulations, the events are random. The crucial period is the initial phase of the simulated epidemic which is decisive for the future evolution of the epidemic, both speed and proportions. As there are very few infecteds the intitial phase of the outbreak proceeds in a highly random fashion. After the initial phase the process smooths out and becomes more predictable and familiar. When considering the effects of changing the distributions it is important to consider effects which befalls the initial phase but are evened out as more people become infected.

The first graph illustrate the impact of the gamma versus the exponential distribution but respresents an endemic scenario. The case of an outbreak is different as rate of people becoming infectious is not constant, but rather grows exponentially. In the second graph, people enter not at a constant rate, but at an exponentially increasing rate such that the rate of entrance every week is ten times what it was the previous week and the tunnel will ever be more and more packed with people. As long as the groth rate does not wane, a steady state is never reached. The slow-walkers will never compensate for the speeders and the number of people in the gamma-tunnel climbs faster than in the exponential tunnel.

In a multi-municipal model the dynamics are more complex and our simplistic cave-model does not offer any enlightenment. The basic behaviour, though, is expected to follow along the same lines as in the single municipality case and the arguments are similar, but to what extent to is not immediately certain. Intuition tells us that the combined effect of two contributions is more than the sum. We may therefore expect a high incidence in when the infectious period is prolonged due to the combined contribution of more numerous infectious and the amount of traveling they have time with during their infectious period. As it turns out, the results of our simulations agrees with preliminary guesses.

We should mention that the gamma-distribution is perhaps not the only choice of modelers. Many alternatives have been proposed such as the Log-normal distribution and Weibull distributions. All three have similar plots but differ some in key points also as regards to the behaviour of the tails. As we have illustrated the tails of the assumed distribution is important for the outcome of the simulations. The effect of these differences for epidemic models have not been studied to our knowledge. None of these, however, would be compatible with our modeling approach which uses stages. In that respect, our choice is as much a consequence of design as deliberate choice, as is the exponential distribution to other model. The significant improvement of the model shown in this paper, while retaining the Markov model. Possible benefits of alternative choices of distributions will be for future experiments to show.

\section{Acknowledgements}

This study was supported by The Swedish Institute for Infectious Diseases Control and the European Union Research NEST Project (DYSONET 012911). The authors would like to express their gratitude to the members of S-GEM, Stockholm Group of Epidemic Modeling http://www.s-gem.se for their kind support.

\bibliography{paper}

\bibliographystyle{paper}

\end{document}